\documentclass[twocolumn,showpacs,preprintnumbers,amsmath,amssymb]{revtex4}
\usepackage{graphicx}% Include figure files
\usepackage{dcolumn}% Align table columns on decimal point
\usepackage{bm}% bold math

\begin{document}

\title{Super-resolution ghost imaging via compressive sampling reconstruction}% Force line breaks with \\

\author{Wenlin Gong}
\email{gongwl@siom.ac.cn}
\author{Shensheng Han}
\email{sshan@mail.shcnc.ac.cn}
\affiliation{ Key Laboratory for
Quantum Optics and Center for Cold Atom Physics of CAS, Shanghai
Institute of Optics and Fine Mechanics, Chinese Academy of Sciences,
Shanghai 201800, China }

\date{\today}% It is always \today, today,

\begin{abstract}
For ghost imaging, pursuing high resolution images and short acquisition times required for reconstructing images are always two main goals. We report an image reconstruction algorithm called compressive sampling (CS) reconstruction to recover ghost images. By CS reconstruction, ghost imaging with both super-resolution and a good signal-to-noise ratio can be obtained via short acquisition times. Both effect influencing and approaches further improving the resolution of ghost images via CS reconstruction, relationship between ghost imaging and CS theory are also discussed.
\end{abstract}

\maketitle
In recent ten years, ghost imaging (GI) has attracted lots of attentions in the field of quantum optics \cite{Gatti1}. The image of
an unknown object can be nonlocally reconstructed by the intensity correlation measurements between two light fields.
Both entangled source and thermal light can be used to realize ghost imaging \cite{Gatti1,Pittman,Angelo,Glauber,Cheng,Gatti2,Bennink,Saleh,Valencia,Gong,Gong1,Zhang,Zhang1,Zhai,Ferri,Gatti3,Bromberg,Katz}.
The researches of ghost imaging have demonstrated that all the information of an object can be obtained by ``global random" measurements and lots of problems which are hard to be solved by conventional imaging approaches, such as x-ray diffraction imaging, projection imaging in Fraunhofer region and imaging in scattering media, can be settled by ghost imaging \cite{Cheng,Gong,Gong1,Zhang}. However, the reconstruction
algorithm of intensity correlation measurements (also called GI reconstruction) always faces with
two main drawbacks. One is that the best resolution of recovered images is determined by the size of the speckle placed on the object
plane based on the previous experimental results \cite{Gong,Ferri}. The other is that long acquisition times should be required for
reconstructing images with a good signal-to-noise ratio (SNR) \cite{Gatti1}.

Recently, compressive sampling (CS) theory has proved mathematically that ``global random" measurements have much higher image extraction efficiency than ``point-by-point" scanning measurements \cite{Candes1,Donoho1,Robucci,Donoho2,Candes2,Romberg}. Correspondingly, an advanced reconstruction algorithm called CS reconstruction has asserted experimentally that one can recover certain signals and images from far fewer samples or measurements than traditional methods use \cite{Katz,Robucci,Romberg,Herman,Figueiredo}. CS reconstruction also offers great potential for better resolution over classical imaging \cite{Herman}. However, in these experiments, the pseudo-thermal incoherent measurement matrix A is obtained by hardware methods and the test detector should collect all the intensity from the object. In practical imaging applications, it is impossible to collect all the intensity from the object in most of the imaging schematics, but the complete image of an object can still be realized by ghost imaging even if the test detector is a pointlike detector \cite{Gong,Gatti3,Zhang}. In this letter, super-resolution ghost imaging via CS reconstruction is investigated when the test detector just collects partial intensity form the object and relationship between ghost imaging and CS theory is discussed.

Fig. 1(a) represents standard schematic of conventional imaging. The experimental schematic for ghost imaging with thermal light is shown in Fig. 1(b). The light source $S$, which is obtained by passing a laser beam through a slowly rotating ground glass disk \cite{Zhang}, first propagates through a beam splitter, then is divided into a test and a reference path. In the test path, the partial intensity which is transmitted through the object is collected by the test detector $D_t$. In the reference path, a copy of the speckle field which impinges on the object is recorded with a CCD camera $D_r$.

\begin{figure}
\centerline{
\includegraphics[width=8.5cm]{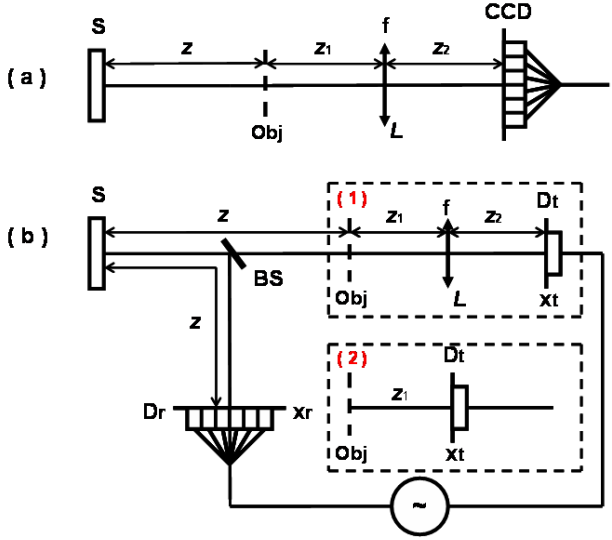}}
\caption{Schematics of conventional imaging and ghost imaging with thermal light.
(a). Conventional imaging; and (b). Ghost imaging; (1) and (2) are the schemes of the test path for the experiment
and the discussion of further improving the resolution of ghost images, respectively.}
\end{figure}

By optical coherence theory \cite{Glauber}, the intensity distribution
obtained by first-order correlation can be represented as:
\begin{eqnarray}
I(x)= \int {dx_1 } \int {dx_2 G^{(1,1)} (x_1 ,x_2)} h_t^*(x,x_1)h_t(x ,x_2).
\end{eqnarray}
where $G^{(1,1)} (x_1 ,x_2)$ is the first-order correlation function on
the source plane. $h_t(x ,x_1)$, $h_t^*(x,x_1)$ are the impulse function of
optical system and phase conjugate of the impulse function, respectively.

Based on ghost imaging via GI reconstruction \cite{Cheng,Gatti2}, we can
obtain the correlation function between the detectors:
\begin{eqnarray}
\Delta G^{(2,2)} (x_r ,x_t ) = \left| \right.\int {dx_1 } \int {dx_2
} G^{(1,1)} (x_1 ,x_2 ) \nonumber\\\times h_r ^ *  (x_1 ,x_r )h_t
(x_2 ,x_t )\left. \right|^2 .
\end{eqnarray}
where $h_t(x_t ,x_2)$ is the impulse function in the
test path whereas $h_r^*(x_r,x_1)$ denotes phase conjugate of the
impulse function in the reference path.

Suppose the light source is fully spatially incoherent, then
\begin{equation}
G^{(1,1)} (x_1 ,x_2 ) = I_0 \delta (x_1  - x_2 ).
\end{equation}
where $I_0$ is a constant, and $\delta(x)$ is Dirac delta function.

For the schematic shown in Fig. 1(b-1), under the paraxial approximation, the impulse
response function of the reference system is
\begin{equation}
h_r (x_r ,x_1 ) \propto \exp \{ \frac{{j\pi }}{{\lambda z }}(x_r - x_1 )^2 \}.
\end{equation}
When $\frac{1}{z_1}+\frac{1}{z_2}=\frac{1}{f}$, then the impulse response function for the test path is
\begin{eqnarray}
h_t (x_t ,x_2 ) \propto \int {dx'} \exp \{ \frac{{j\pi }}{{\lambda z}}(x' - x_2 )^2 \} t(x')\nonumber\\\times\exp \{ \frac{{j\pi }}{{2\lambda z_1}}x'^2\}\sin c[\frac{L}{\lambda}(\frac{x_t}{z_2} + \frac{ x'}{z_1})]\}.
\end{eqnarray}
where t(x), $L$ are the transmission function of the object and effective aperture of the imaging lens $f$, respectively. Furthermore, $\sin c(x)=\frac{\sin(\pi x)}{\pi x}$. Substituting Eqs. (3) and (5) into Eq. (1), the intensity distribution on the CCD camera in Fig. 1(a) is
\begin{equation}
I(x) \propto \int {dx'} \left| {t(x')} \right|^2 \sin c^2 [\frac{L}{{\lambda}}(\frac{x}{z_2} + \frac{ x'}{z_1})].
\end{equation}

Similarly, substituting Eqs. (3)-(5) into Eq. (2), the correlation function for ghost imaging can be represented as
\begin{eqnarray}
\Delta G^{(2,2)} (x_r ,x_t ) \propto \left| {\int {dx'} \sin c[\frac{D}{{\lambda z}}(x_r  - x')]t(x')} \right.\nonumber\\\times \left. {\exp \{ \frac{{j\pi }}{{2\lambda f}}x'^2\}\sin c[\frac{L}{{2\lambda f}}(x_t  + x')]} \right|^2.
\end{eqnarray}
where $D$ is the transverse size of the source. If the smallest length scale of the object is larger than the size of the diffraction limit cased by the lens' effective aperture $L$ (namely the lens $f$ can nearly collect all the intensity from the object), thus $\sin c[\frac{L}{{\lambda }}(\frac{x_t}{z_2}  + \frac{x'}{z_1})]\sim \delta(\frac{x_t}{z_2}  + \frac{x'}{z_1})$, then
\begin{eqnarray}
\Delta G^{(2,2)} (x_r ) = \int {dx_t \Delta G^{(2,2)} (x_r ,x_t )}\nonumber\\\sim \int {dx'} \left| {t(x')} \right|^2 \sin c^2 [\frac{D}{{\lambda z}}(x_r  - x')].
\end{eqnarray}

From Eqs. (6) and (8), the best resolution of images are determined by the size of the speckle placed on the object plane (namely $\Delta x \sim \frac{{\lambda z}}{D}$) for ghost imaging via GI reconstruction whereas for conventional imaging depend on the effective aperture of the imaging lens $f$ (namely $\Delta x \sim \frac{{\lambda z_1}}{L}$).

However, for ghost imaging, except for GI reconstruction to recover the object's transmission function, another novel method called CS reconstruction algorithm perfectly accords with the physical principle of ghost imaging and can be used to recover the object. By CS theory \cite{Candes1,Donoho1,Robucci,Donoho2,Candes2,Romberg,Herman,Figueiredo}, when the image to be recovered with $n\times n$ matrix can be viewed as a vector in $R^{n^2}$, the measurements process can be written compactly in matrix notation as
\begin{equation}
y=Ax.
\end{equation}
where $x\in R^{n^2}$ is the ``true" image. The $m\times R^{n^2}$ matrix $A$ is constructed by stacking the $m$ measurement basis functions $\longrightarrow$ each of which is also a vector in $R^{n^2}$ and any two of which are mutually incoherent. $y$ is the $m$ vector containing the observations. Given $y$, we reconstruct the image by solving the following convex optimization program \cite{Figueiredo}:
\begin{eqnarray}
\mathop {\min }\limits_x \left\| x \right\|_{\ell_1 }\ {\rm{subject \ to}} \ \ y = Ax.
\end{eqnarray}
where $\left\| V \right\|_{ \ell_1 }  = \sum\nolimits_i {\left| {v_i } \right|}$ is the $\ell_1$ norm of $V$.

In the schematic of ghost imaging shown in Fig. 1(b), the $m\times R^{n^2}$ matrix $A$ can be obtained by the reference path and each of the measurement basis functions is constructed by one of realizations recorded with the CCD camera $D_r$. Because each of the speckle fields registered by the CCD camera $D_r$ is random and independent of others, thus the property of incoherence sampling among the $m$ vectors of matrix $A$ is obviously satisfied. Similarly, the $m$ vector $y$ is corresponded to the $m$ measurements registered by the bucket detector $D_t$. If the speckle field which impinges on the object is described by $I_r(x,y)$ and $B_r$ denotes the total intensity recorded by the detector $D_t$, then the image of the object can be reconstructed by solving the following convex optimization program by Eq. (10):
\begin{eqnarray}
t_{CS}  = t' \ {\rm{which \ minimizes}}: \ \left\| t'(x,y) \right\|_{l_1 }; \ {\rm{subject \ to}} \nonumber\\\int {dx} \int {dy} I_r (x,y)t'(x,y) = B_r ,\forall _r  = 1 \cdots m.
\end{eqnarray}
where $t_{CS}$ is the object's transmission function recovered by CS reconstruction algorithm.

For CS reconstruction, the resolution of reconstructed object depends on the row vector of matrix A (namely $R^{n^2}$), while the image of an object is obtained by scanning the position of the photons on the CCD camera $D_r$ for ghost imaging. Thus the resolution of ghost imaging recovered via CS reconstruction is closely related to the resolution of the CCD camera $D_r$ to record $I_r(x,y)$.

In the experiment, the wavelength of the source was $\lambda$=650nm and the transverse size of the source was $D$=2.0mm. Fig. 2 presents the numerical simulation results for conventional imaging, ghost imaging via GI and CS reconstructions when the test path of ghost imaging is proposed as Fig. 1(b-1). The object is a double-slit with slit width $a$=30$\mu$m, slit height $h$=120$\mu$m and center-to-center separation $d$=60$\mu$m. For the CS reconstruction, we have utilized the $\ell_1$-Magic algorithm \cite{Donoho1,Figueiredo}. From Eqs. (6) and (8), the resolution of both conventional imaging and ghost imaging recovered by GI reconstruction are very low, which are also demonstrated in Fig. 2(b-c). However, a high-resolution ghost image can be obtained by CS reconstruction (Fig. 2(d)). Correspondingly, from Fig. 2(d-e), the resolution of recovered images via CS reconstruction will reduce as the resolution of the CCD cameras $D_r$ is decreased.

\begin{figure}
\centerline{
\includegraphics[width=8.5cm]{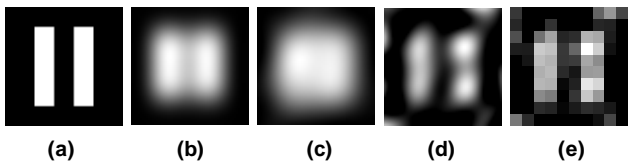}}
\caption{Simulated conventional imaging, ghost imaging via GI and CS reconstructions of a double-slit with $z$=200mm, $f$=250mm, $z_1$=$z_2$=500mm and $L$=6mm when the test path of ghost imaging is proposed as Fig. 1(b-1). (a). The object; (b). Conventional imaging; (c). Ghost imaging via GI reconstruction with 2000 realizations; (d) and (e) are ghost imaging via CS reconstruction with 32 realizations when the pixel resolution of the CCD camera $D_r$ is 3$\mu$m and 18$\mu$m, respectively.}
\end{figure}

For ghost imaging, the lens $f$ shown in Fig. 1(b-1) is just used to collect the information from the object and the lens' effective aperture $L$ limits the information of high frequency. In practical applications, we can use a device with large collecting area to collect the information from the object instead of the lens $f$, which is similar to the case in Fig. 1(b-2). In Fig. 1(b), by replacing the test path (1) using the scheme (2), ghost imaging via GI and CS reconstructions of the same double-slit in different collecting areas for the test detector $D_t$ are shown in Fig. 3. From Fig. 3(a-b), enlarging the collecting areas of the test detector can improve the resolution of ghost imaging when $a<\frac{{1.22\lambda z_1}}{L_1}$(namely the collecting areas of the test detector is smaller than 13.2mm$\times$13.2mm). However, similar to the results described in Eq. (8), if $a>\frac{{1.22\lambda z_1}}{L_1}$, then the best resolution of ghost imaging via GI reconstruction will be determined only by the size of the speckle placed on the object plane, which is also been demonstrated in Fig. 3(b). Correspondingly, more information from the object is collected by the test detector because of the increase of the collecting areas, thus the quality of ghost imaging via CS reconstruction will be enhanced (Fig. 3 (c-d)). Fig. 3(c-d) also shows that the resolution of recovered images via CS reconstruction will reduce as the decrease of the resolution of the CCD cameras $D_r$.
\begin{figure}
\centerline{
\includegraphics[width=8.5cm]{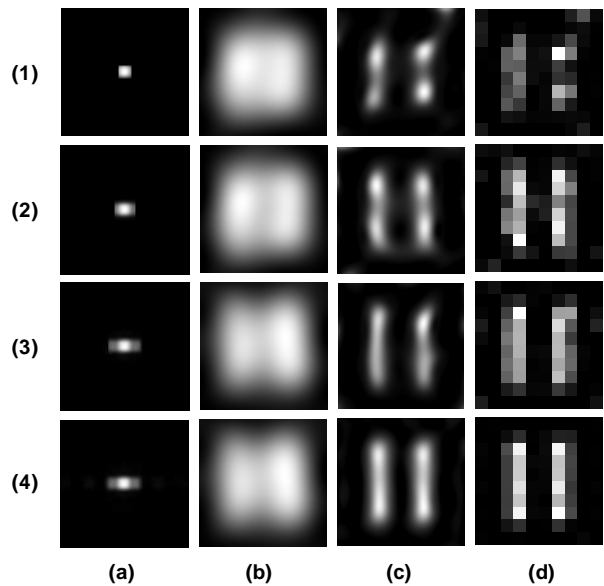}}
\caption{Numerical simulation results of ghost imaging via GI and CS reconstructions with $z_1$=500mm when Fig. 1(b-2) is used as the test path of ghost imaging. (a). The intensity distribution on the test detector plane with 2000 realizations; (b). GI reconstruction with 2000 realizations; (c) and (d), respectively, are CS reconstruction with 32 realizations when the pixel resolution of the CCD camera $D_r$ is 3$\mu$m and 18$\mu$m. The collecting areas of the test detector $D_t$ with the sizes $L_1\times L_1$ shown in (1), (2), (3) and (4) are 6mm$\times$6mm, 10mm$\times$10mm, 15mm$\times$15mm and 30mm$\times$30mm, respectively.}
\end{figure}

For conventional imaging shown in Fig. 1(a), the image registered by the CCD camera is obtained by long exposure. So the intensity distribution of the light field on the object plane is uniform and CS reconstruction can not be applied because of coherence sampling. Different from conventional imaging, ghost imaging is based on the principle of short exposure and the acquisitions recorded with the CCD cameras $D_r$ are incoherence sampling. Thus, the process of ghost imaging shown in Fig. 1(b) describes vividly the standard CS theory but we also demonstrated that ghost imaging can be recovered by CS reconstruction even if the intensity from the object are partially collected.

Furthermore, for conventional imaging, enlarging the effective aperture of imaging lens $f$, the resolution of the image can be improved. Also, the resolution of ghost images via GI reconstruction can be enhanced by decreasing the transverse coherence width placed on the object plane (such as a source with large transverse size, a small distance $z$). However, the resolution of recovered images via CS reconstruction is related to the resolution of the CCD camera $D_r$ (Fig. 2(d-e), Fig. 3(c-d)). So for ghost imaging via CS reconstruction, the resolution of ghost images can be further improved by increasing the resolution of the CCD camera $D_r$ or computational ghost imaging approaches. From Fig. 3(a) and (c-d), enlarging the collecting areas of the test detector can also enhance the quality of the images recovered by CS reconstruction. Fig. 2(b-d) and Fig. 3(b-c) also shows that CS reconstruction requires much shorter acquisition times than GI reconstruction for reconstructing ghost images with a good SNR, which have already been demonstrated in Ref. \cite{Katz}.

In conclusion, ghost imaging via CS reconstruction is the result combining quantum optics with information theory. Based on CS reconstruction, except for allowing for shorter acquisition times for reconstructing ghost images with a good SNR, super-resolution ghost images can also be obtained even if the information from the object is partially collected and the best resolution of reconstructed images is closely related to the CCD camera registering the light field in the reference path. In practical applications, enlarging the collecting areas of the test detector is much easier than making a lens with large effective aperture. So ghost imaging via CS reconstruction is very useful to the imaging in the long wavelength radiation band and in the far field, microscopy, astronomy and so on.

The work was partly supported by the Hi-Tech Research and
Development Program of China under Grant Project No. 2006AA12Z115,
National Natural Science Foundation of China under Grant Project No.
60877009, and Shanghai Natural Science Foundation under Grant Project No. 09JC1415000.

\end{document}